\documentclass[amsmath,aps,prl,reprint,twocolumn]{revtex4-1}

\usepackage{graphicx,hyperref}
\usepackage[varg]{txfonts}

\begin{document}

\newcommand{\D}{\mathrm{d}}
\newcommand{\E}{\mathrm{e}}
\newcommand{\I}{\mathrm{i}}
\newcommand{\cvector}[1]{\left(\begin{array}{c}#1\end{array}\right)}
\newcommand{\tens}[1]{\boldsymbol{#1}}

\title{Analytic calculation of the non-linear cosmic density-fluctuation power spectrum in the Born approximation}
\date{\today}
\author{Matthias Bartelmann}
\author{Felix Fabis}
\author{Sara Konrad}
\author{Elena Kozlikin}
\author{Robert Lilow}
\author{Carsten Littek}
\author{Johannes Dombrowski}
\affiliation{Universit\"at Heidelberg, Zentrum f\"ur Astronomie, Institut f\"ur Theoretische Astrophysik}

\begin{abstract}
  We derive a non-perturbative, closed, analytic equation for the non-linear power spectrum of cosmic density fluctuations. This result is based on our kinetic field theory (KFT) of cosmic structure formation and on evaluating particle interactions in the Born approximation. At redshift zero, relative deviations between our analytic result and typical numerical results are $\sim15\%$ on average up to wave numbers of $k\le10\,h\,\mathrm{Mpc}^{-1}$. The theory underlying our analytic equation is fully specified once the statistical properties of the initial density and momentum fluctuations are set. It has no further adjustable parameters. Apart from this equation, our main result is that the characteristic non-linear deformations of the power spectrum at late cosmic times are determined by the initial momentum correlations and a partial compensation between diffusion damping and particle interactions.
\end{abstract}


\maketitle

\section{Introduction}

Based on the Martin-Siggia-Rose formalism \cite{1973PhRvA...8..423M} for the statistical dynamics of classical systems and on the pioneering work by Mazenko \cite{2010PhRvE..81f1102M, 2011PhRvE..83d1125M} and Das and Mazenko \cite{2012JSP...149..643D, 2013JSP...152..159D}, we have developed a microscopic, non-equilibrium kinetic field theory (KFT) for cosmic structure formation \cite{2016NJPh...18d3020B}. This theory dissolves the cosmic dark-matter density field into classical particles obeying Hamiltonian dynamics. Once the initial probability distribution of the particles in phase space has been specified, including correlations between all phase-space coordinates $x = (q, p)$, the statistical properties of the particle ensemble are incorporated into a generating functional whose time evolution is determined by the Green's function of the Hamiltonian equations of motion. Retaining the full phase-space information avoids the notorious shell-crossing problem. The Hamiltonian flow in phase-space establishes a diffeomorphic map of the stochastic initial conditions to the statistical properties of the particle ensemble at later times.

Based on KFT and on a suitable choice of the Green's function \cite{2015PhRvD..91h3524B}, we showed in \cite{2016NJPh...18d3020B} that first-order perturbation theory for the particle interactions reproduces quite well the non-linear evolution of the cosmic density-fluctuation power spectrum as found in numerical simulations. In that paper, we furthermore approximated the initial momentum correlations up to second order in the momentum-correlation function. In \cite{2017NJPh...19h3001B}, we showed how the free generating functional of KFT can be completely factorized if the full hierarchy of initial momentum correlations is taken into account.

Here, we avoid the perturbative approach to the particle interactions and instead include a suitably averaged interaction term into the generating functional of KFT. This allows us to derive a closed, analytic, non-perturbative equation for the non-linear cosmic density-fluctuation power spectrum which reproduces numerical results for late cosmic times very well. The theory is free of adjustable parameters. The only freedom it allows lies in the initial conditions, the interaction potential and the Green's function of the equations of motion. We show that the characteristic deformation of the non-linear compared to the linear power spectrum is due to a partial compensation between momentum diffusion and particle interaction and to the initial hierarchy of momentum correlations.

\section{Particle interactions in the generating functional}

We only briefly review KFT here and refer the interested readers to \cite{2016NJPh...18d3020B, 2017NJPh...19h3001B}. Following \cite[Eq.~31]{2016NJPh...18d3020B}, the complete generating functional of KFT is
\begin{equation}
  Z[\tens J] = \int\D\Gamma\exp\left[
    \I\int_0^\infty\left\langle\tens J(t'), \tens x(t')\right\rangle\D t'
  \right]
\label{eq:1}
\end{equation}
where the angular brackets in the phase factor denote a scalar product between the generator field $\tens J$ and the set of $N$ particle trajectories in phase space $\tens x(t')$, determined by the Green's function of the equations of motion. The initial distribution $P(\tens q, \tens p)$ of the phase-space coordinates is incorporated in the integral measure $\D\Gamma = P(\tens q, \tens p)\D\tens q\D\tens p$, where $(q, p)$ without a time argument are the initial phase-space coordinates.

By applying two one-particle density operators representing density modes with wave vectors $\vec k_{1,2}$ at time $t$ to $Z$ and setting $\tens J=0$ afterwards, the generating functional returns the density power spectrum $G_{\rho\rho}(12)$,
\begin{equation}
  Z[\tens L] = \mathcal{N}G_{\rho\rho}(12) = \int\D\Gamma\,
  \E^{\I\langle\tens L_q, \tens q\rangle+
      \I\langle\tens L_p(t), \tens p\rangle-\bar F(t)}\;,
\label{eq:5}
\end{equation}
where $\mathcal{N}$ is a normalizing prefactor which is irrelevant for our purposes. The shift tensors
\begin{equation}
  \tens L_q := -\vec k_1\otimes\left(\vec e_1-\vec e_2\right)\;,\quad
  \tens L_p(t) := g_{qp}(t,0)\,\tens L_q
\label{eq:4}
\end{equation}
are introduced by the density operators, with statistical homogeneity enforcing $\vec k_1 = -\vec k_2$. The propagator $g_{qp}(t,t')$ will be specified in (\ref{eq:32}) below. The time-integrated interaction term
\begin{equation}
  \bar F(t) =
  \I\,\int_0^t\D t'\langle\tens L_p(t'),\nabla\tens V(t')\rangle =:
  \int_0^t\D t'F(t,t')
\label{eq:6}
\end{equation}
contains the interaction potential $\tens V$ between the particles. We emphasize that the expressions (\ref{eq:1}) for the generating functional and (\ref{eq:6}) for the interaction term are exact.

\section{Particle interactions in the Born approximation}

We will now evaluate the interaction term (\ref{eq:6}) in the Born approximation for the particle trajectories and average it over particle positions. The potential gradient at time $t'$ between two arbitrary particles $1$ and $2$ is
\begin{equation}
  \nabla_1V_2(t') = \I\int_k\vec k\,\tilde v\left(k,t'\right)\,
  \E^{\I\vec k\cdot\left[\vec q_1(t')-\vec q_2(t')\right]}\;,
\label{eq:9}
\end{equation}
where $\tilde v$ is the Fourier transform of the two-particle interaction potential. The particle coordinates are to be taken at the time $t'$ of the interaction and are thus not to be confused with their initial values. In this expression, we replace the actual particle trajectories $\vec q(t')$ by the unperturbed, \emph{inertial} trajectories and set
\begin{equation}
  \vec q(t') \to \vec q+g_{qp}(t',0)\,\vec p\;,
\label{eq:12}
\end{equation} 
where $\vec q$ and $\vec p$ without a time argument are initial particle positions and momenta, respectively. 

Now, we average the phase factor in (\ref{eq:9}) by integrating it over the phase-space distribution of the particles appearing in the integrand of (\ref{eq:5}), but setting the interaction term $\bar F(t)$ to zero. This means that we evaluate the phase-space distribution of the particles at the shifts (\ref{eq:4}) corresponding to the density modes involved, and evolved to time $t'$ with the Born approximation. Accordingly, we carry out the normalized integral
\begin{align}
  \left\langle
    \E^{\I\vec k\cdot\left[\vec q_1(t')-\vec q_2(t')\right]}
  \right\rangle = \frac{1}{\mathcal{N}}\int\D\Gamma
    \E^{\I\langle\tens L_q,\tens q\rangle+
        \I\langle\tens L_p(t'),\tens p\rangle+
        \I\vec k\cdot\left[\vec q_1(t')-\vec q_2(t')\right]}\;,
\label{eq:13}
\end{align}
with the shift $\tens L_p(t')$ from (\ref{eq:4}) evaluated at $t'$. This shows that the average phase factor can be brought into the form
\begin{equation}
  \left\langle
    \E^{\I\vec k\cdot\left[\vec q_1(t')-\vec q_2(t')\right]}
  \right\rangle = \frac{1}{\mathcal{N}}\int\D\Gamma
  \E^{\I\langle\tens L'_q,\tens q\rangle+
      \I\langle\tens L'_p(t'),\tens p\rangle}\;,
\label{eq:15}
\end{equation}
where now the modified shifts
\begin{equation}
  \tens L_q' := -\vec\kappa\otimes
    \left(\vec e_1-\vec e_2\right)\;,\quad
  \tens L_p'(t') := g_{qp}(t',0)\tens L_q'
\label{eq:16}
\end{equation}
denoted with a prime occur, where $\vec\kappa := \vec k_1-\vec k$.

The phase-space integral remaining in (\ref{eq:15}) is the free generating functional $Z_0$, evaluated at the modified shift tensor $\tens L'$ with components given in (\ref{eq:16}),
\begin{equation}
  \int\D\Gamma
  \E^{\I\langle\tens L'_q,\tens q\rangle+
      \I\langle\tens L'_p(t'),\tens p\rangle} =
  Z_0[\tens L']\;.
\label{eq:17}
\end{equation} 
As shown in \cite[Eq.~42]{2017NJPh...19h3001B},
\begin{equation}
  Z_0[\tens L'] = \mathcal{N}\E^{Q_\mathrm{D}(\kappa, t')}\left[
    (2\pi)^3\delta_\mathrm{D}(\vec\kappa\,)+\mathcal{P}(\kappa, t')
  \right]
\label{eq:18}
\end{equation}
with
\begin{equation}
  \mathcal{P}(\kappa, t') =
  \int_q\left\{
    \E^{-g_{qp}^2(t',0)\kappa^2a_\parallel(q)}-1
  \right\}\E^{\I\vec\kappa\cdot\vec q}\;.
\label{eq:19}
\end{equation}
The function $a_\parallel(q)$ appearing here is the correlation function of the momentum components parallel to the line connecting the correlated particles,
\begin{equation}
  a_\parallel(q) = \mu^2\xi_\psi''(q)+(1-\mu^2)\frac{\xi_\psi'(q)}{q}\;,
\label{eq:19a}
\end{equation}
(cf.~\cite[Eq.~B.28]{2017NJPh...19h3001B}), where $\xi_\psi(q)$ is the correlation function of the initial velocity potential and $\mu$ is the angle cosine between the vectors $\vec\kappa$ and $\vec q$.

The momentum-diffusion term $Q_\mathrm{D}(\kappa, t')$ appearing in (\ref{eq:18}) is
\begin{equation}
  Q_\mathrm{D}(\kappa, t') = -\frac{\sigma_1^2}{3}g_{qp}^2(t',0)\kappa^2\;.
\label{eq:21}
\end{equation}
As shown in \cite[Eq.~B.41]{2017NJPh...19h3001B}, we can approximate the evolved power spectrum $\mathcal{P}(\kappa, t')$ from (\ref{eq:19}) by the linearly evolved density-fluctuation power spectrum,
\begin{equation}
  \mathcal{P}(\kappa, t') \approx g_{qp}^2(t',0)P_\delta(\kappa)\;,
\label{eq:20}
\end{equation}
for sufficiently small arguments of the exponential in (\ref{eq:19}). With these results, we can write the average phase factor from (\ref{eq:15}) as
\begin{equation}
  \left\langle
    \E^{\I\vec k\cdot\left[\vec q_1(t')-\vec q_2(t')\right]}
  \right\rangle = (2\pi)^3\delta_\mathrm{D}\left(\vec\kappa\,\right)+
  \bar P_\delta\left(\kappa, t'\right)
\label{eq:22}
\end{equation}
with the damped, evolved density-fluctuation power spectrum $\bar P_\delta(\kappa, t') := \E^{Q_\mathrm{D}(\kappa,t')}\,g_{qp}^2(t',0)\,P_\delta(\kappa)$. For a two-point function, only $\nabla_1V_2$ and $\nabla_2V_1 = -\nabla_1V_2$ remain after averaging over the particle distribution. The averaged interaction term in the Born approximation thus reads
\begin{align}
  \langle F(t,t')\rangle &= 2g_{qp}(t,t') \nonumber\\
  &\cdot\left[
    k_1^2\tilde v\left(k_1,t'\right)+
    \vec k_1\cdot\int_k\,\vec k\,\tilde v\left(k,t'\right)
    \bar P_\delta\left(\kappa, t'\right)
  \right]\;.
\label{eq:11}
\end{align}

To proceed, we specialize the Fourier transform of the two-particle interaction potential to
\begin{equation}
  \tilde v(k, t') = A(t')k^{-2}\;,
\label{eq:23}
\end{equation}
i.e.\ to the Newtonian spatial dependence. The amplitude to be specified below depends on time because of cosmic expansion. Then, we can write
\begin{align}
  \left\langle F(t, t')\right\rangle &=
  2\,g_{qp}(t,t')\,A(t')\left[
    1+\int_k\frac{\vec k_1\cdot\vec k}{k^2}\,
    \bar P_\delta(\kappa, t')
  \right]\;.
\label{eq:24}
\end{align}
The remaining integral over the evolved power spectrum can be further simplified by transforming the integration variable to $\vec\kappa$, introducing $y := \kappa/k_1$ and the angle cosine $\mu$ between $\vec\kappa$ and $\vec k_1$. Then,
\begin{equation}
  \int_k\frac{\vec k_1\cdot\vec k}{k^2}\,\bar P_\delta(\kappa, t') =
  k_1^3\int_0^\infty\frac{y^2\D y}{(2\pi)^2}\,
  \bar P_\delta\left(k_1y, t'\right)J(y)
\label{eq:25}
\end{equation}
with the function $J(y)$ defined by
\begin{equation}
  J(y) := \int_{-1}^1\D\mu\,\frac{1-y\mu}{1+y^2-2y\mu} =
  1+\frac{1-y^2}{2y}\ln\frac{1+y}{|1-y\,|}\;.
\label{eq:26}
\end{equation}

\section{Specification of the interaction potential}

We have shown in \cite{2015PhRvD..91h3524B} that the Hamiltonian in an expanding space-time is given by
\begin{equation}
  \mathcal{H} = \frac{\vec p^{\,2}(t')}{2g(t')}+gV\;,
\label{eq:28}
\end{equation}
if the linear growth factor $D_+$ is used as a time coordinate, $t := D_+-D_+^\mathrm{(i)}$. The time-dependent function $g(t')$ is normalized such that $g(0) = 1$ at the initial time. Since all particles are assumed to have the same mass, we may set $m = 1$. The potential $V$ satisfies the Poisson equation
\begin{equation}
  \nabla^2V = \frac{3}{2}\frac{a}{g^2}\,\delta\;,
\label{eq:29}
\end{equation}
where $a$ is the usual cosmological scale factor and $\delta$ is the density contrast. The Hamiltonian (\ref{eq:28}) implies the equation of motion
\begin{equation}
  \ddot{\vec q}(t') = -\frac{\dot g}{g}\,\dot{\vec q}(t')-\nabla V\;.
\label{eq:31}
\end{equation}
Note that, since $g = 1$ initially, $\dot{\vec q} = \vec p$ at the initial time. We wish to express the solution of (\ref{eq:31}) by a propagator $g_{qp}(t,t')$ such that
\begin{equation}
  \vec q(t') = \vec q+g_{qp}(t',0)\vec p+
  \int_0^{t'}\D\bar t\,g_{qp}(t',\bar t)\vec f(\bar t)\;,
\label{eq:32}
\end{equation}
with a time-dependent force kernel $\vec f(\bar t)$ to be determined. Note that $\vec f(\bar t)$ then assumes the role of $-\nabla V$ introduced in (\ref{eq:6}). The second time derivative of (\ref{eq:32}) is
\begin{equation}
  \ddot{\vec q}(t') = \ddot g_{qp}(t',0)\vec p+
  \dot g_{qp}(t',t')\vec f(t')+\int_0^{t'}\D\bar t\,
  \ddot g_{qp}(t',\bar t)\vec f(\bar t)\;.
\label{eq:33}
\end{equation} 
The kernel $\vec f(t')$ now needs to be determined such that, given $g_{qp}(t,t')$, (\ref{eq:33}) reproduces (\ref{eq:31}).

For the Zel'dovich approximation, the propagator is simply $g_{qp}(t,t') = t-t'$. Inserting this into (\ref{eq:33}) leads to $\ddot{\vec q}(t') = \vec f(t')$ which implies, with (\ref{eq:31}),
\begin{equation}
  \vec f(t') = -\frac{\dot g}{g}\,\dot{\vec q}(t')-\nabla V\;.
\label{eq:34}
\end{equation}
Using the Zel'dovich propagator thus requires us to augment the effective force term by a potential-dependent term in addition to the negative potential gradient.

In \cite{2015PhRvD..91h3524B}, we have proposed the improved Zel'dovich propagator
\begin{equation}
  g_{qp}(t,t') = \int_{t'}^t\D\bar t\,
    \E^{h(\bar t)-h(t')}
  \quad\mbox{with}\quad h(t) := \frac{1}{g(t)}-1
\label{eq:35}
\end{equation}
such that $h(0) = 0$. With
\begin{equation}
  \dot  g_{qp}(t,t') = \E^{h(t)-h(t')}\;,\quad
  \ddot g_{qp}(t,t') = \dot h(t)\,\E^{h(t)-h(t')}
\label{eq:36}
\end{equation}
and (\ref{eq:33}), this implies
\begin{equation}
  \ddot{\vec q}(t') = \dot h(t')\,\dot{\vec q}(t')+\vec f(t')\;.
\label{eq:37}
\end{equation}
Equating the result (\ref{eq:37}) for $\ddot{\vec q}(t')$ to (\ref{eq:31}) immediately gives
\begin{equation}
  \vec f(t') =
  -\frac{\dot g}{g}\left(1-\frac{1}{g}\right)\dot{\vec q}(t')-\nabla V\;.
\label{eq:38}
\end{equation}

While the potential-gradient term in (\ref{eq:38}) correlates particle positions, the velocity term correlates the velocity of one particle with the position of another. According to (\ref{eq:12}), the correlation function of particle positions in the Born approximation grows proportional to $g_{qp}^2(t',0)$, while the cross-correlation function between particle velocities and positions evolves proportional to $\dot g_{qp}(t',0)g_{qp}(t',0)$. The velocity-dependent term in (\ref{eq:38}) thus grows with a rate lowered by the ratio $\dot g_{qp}(t',0)/g_{qp}(t',0)$ compared to the potential-gradient term. Apart from that, both terms in (\ref{eq:38}) depend in the same way on the wave vector $\vec k$ in Fourier space, i.e.\ proportional to $\I\vec k\,k^{-2}\,P_\delta(k)$, where the factor $\I\vec k$ is due to the gradient of the initial velocity potential in the first case, and to the gradient of the interaction potential in the second. Thus, the velocity contribution to the force term can be taken into account by writing the Fourier-transformed potential in (\ref{eq:23}) as
\begin{equation}
  \tilde v(k, t') \to \left(A_1(t')+A_2(t')\right)\,k^{-2}
\label{eq:39}
\end{equation} 
with the time-dependent amplitudes
\begin{equation}
  A_1(t') := -\frac{\dot g}{g}
    \left(1-\frac{1}{g}\right)\frac{\dot g_{qp}(t',0)}{g_{qp}(t',0)}\;,\quad
  A_2(t') := -\frac{3}{2}\frac{a}{g^2}\;.
\label{eq:40}
\end{equation}
We abbreviate $A(t') := A_1(t')+A_2(t')$ in the following.

\section{Results for the evolved power spectrum}

\begin{figure}[ht]
  \includegraphics[width=\hsize]{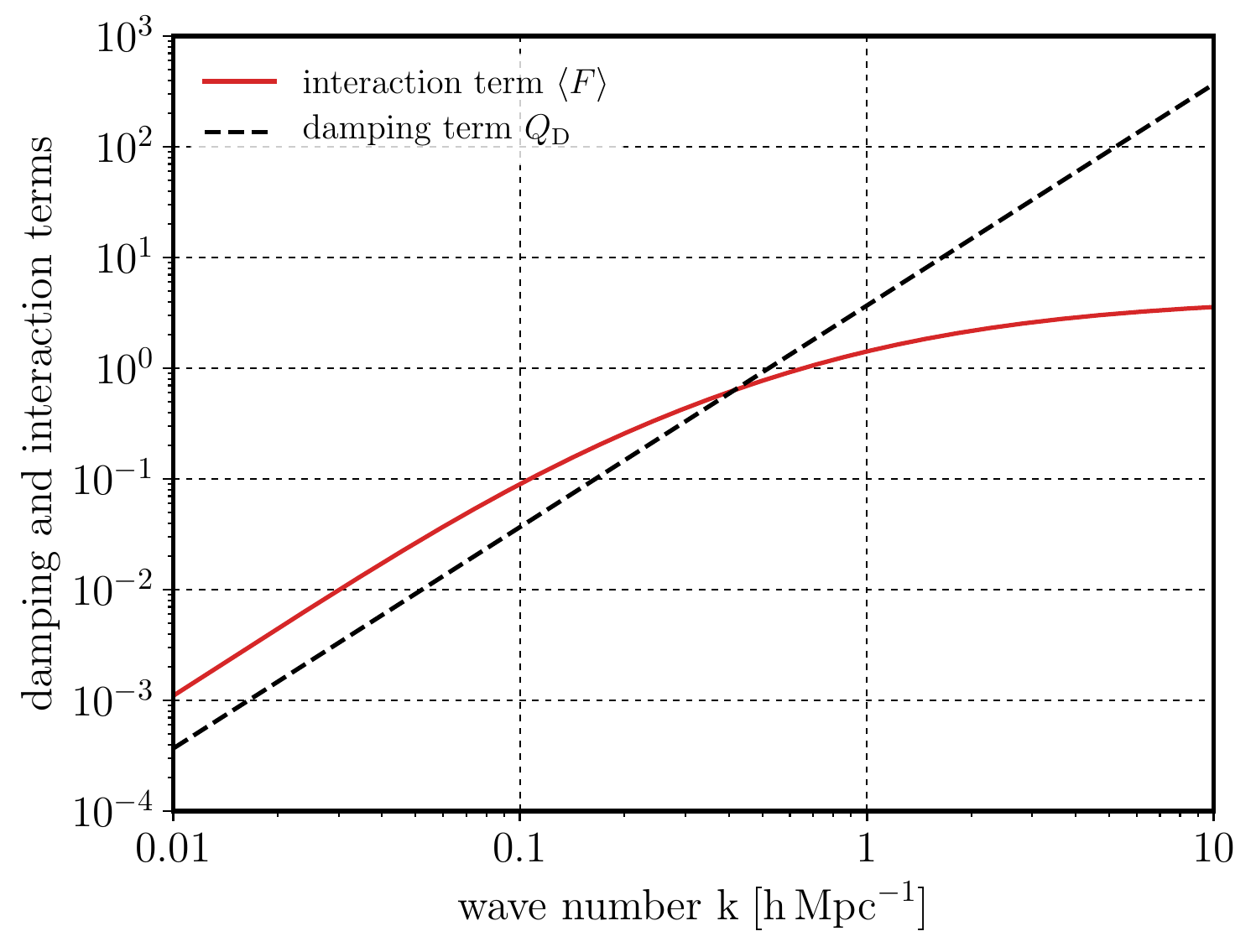}
\caption{Averaged, time-integrated interaction term $\langle\bar F(t)\rangle$ as a function of wave number $k$, evaluated using the Born approximation. Also shown is the momentum-diffusion term $Q_\mathrm{D}(k,t)$. Both curves are shown for the present cosmic time, i.e.\ for $a = 1$.}
\label{fig:1}
\end{figure}

We thus arrive at the result that the averaged interaction term reads
\begin{align}
  \left\langle F(t,t')\right\rangle &=
  2\,g_{qp}(t,t')\,A(t')\nonumber\\ &\cdot \left[
    1+k_1^3\int_0^\infty\frac{y^2\D y}{(2\pi)^2}\,\bar P_\delta(k_1y,t')J(y)
  \right]\;.
\label{eq:41}
\end{align}
We integrate it over time according to (\ref{eq:6}) and replace the interaction term $\bar F(t)$ in the generating functional (\ref{eq:5}) by the result
\begin{equation}
  \langle\bar F(t)\rangle = \int_0^t\D t'\langle F(t,t')\rangle\;.
\label{eq:42}
\end{equation}
This averaged, time-integrated interaction term is shown together with the momentum-diffusion term $Q_\mathrm{D}$ in Fig.~\ref{fig:1} for the present cosmic time. The figure shows that the diffusion term exceeds the interaction term on small scales, $k\gtrsim0.4\,h\,\mathrm{Mpc}^{-1}$. The particle interaction can thus only partially compensate the momentum diffusion.

\begin{figure}[ht]
  \includegraphics[width=\hsize]{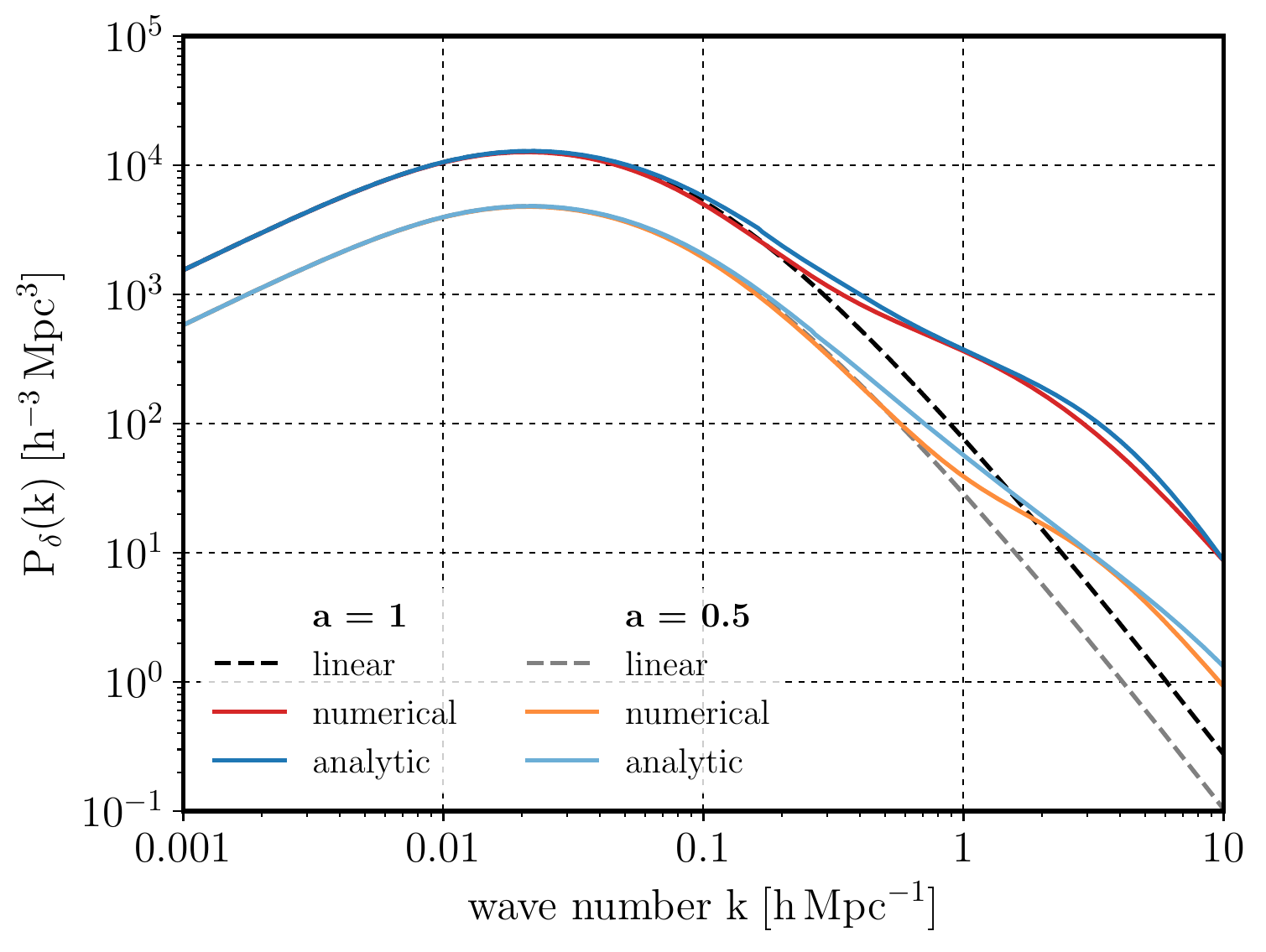}
\caption{Numerical and analytic non-linear power spectra for $a = 0.5$ and $a = 1$. The dashed curves are the input power spectrum, linearly evolved to the same scale factors. The numerical results are modelled following \cite{2003MNRAS.341.1311S}.}
\label{fig:2}
\end{figure}

Since the resulting exponential factor does not depend on the phase-space coordinates, it can be pulled in front of the integral. This allows us to write the evolved power spectrum including the averaged interaction term as
\begin{equation}
  G_{\rho\rho}(k, t) = \E^{Q_\mathrm{D}(k,t)-\langle\bar F(t)\rangle}\left[
    (2\pi)^3\delta_\mathrm{D}(k)+\mathcal{P}(k, t)
  \right]\;.
\label{eq:42}
\end{equation}
The delta distribution is unimportant since it sets $k$ to zero and thus corresponds to a constant mean density. We thus find that the evolved, non-linear density-fluctuation power spectrum is
\begin{equation}
  \bar{\mathcal{P}}(k,t) =
  \E^{Q_\mathrm{D}(k,t)-\langle\bar F(t)\rangle}\int_q\left\{
    \E^{-g_{qp}^2(t)a_\parallel(q)k^2}-1
  \right\}\E^{\I\vec q\cdot\vec k}\;.
\label{eq:43}
\end{equation}
This result is shown in Fig.~\ref{fig:2} for final times or scale factors $a = 1$ and $a = 0.5$.

\begin{figure}[ht]
  \includegraphics[width=\hsize]{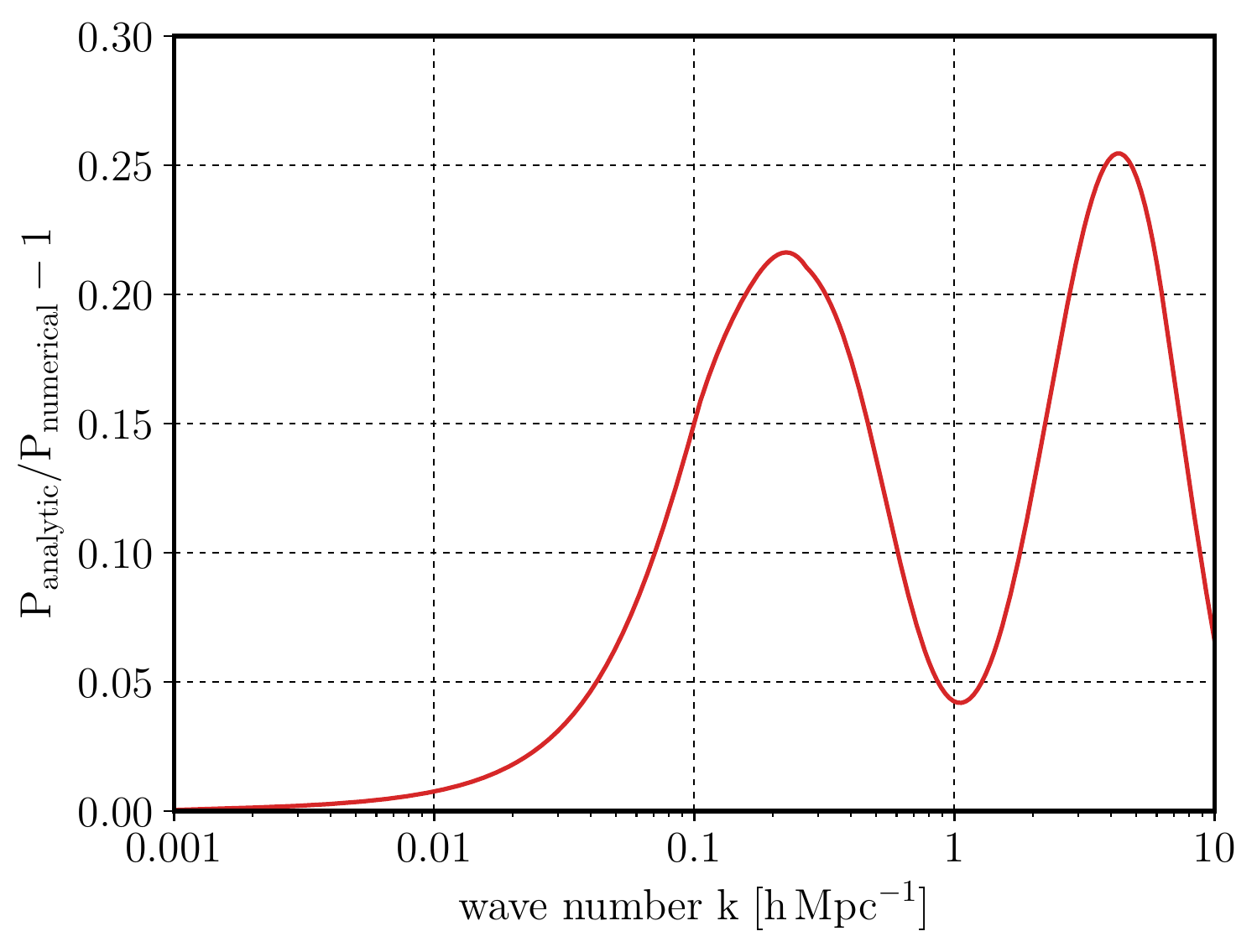}
\caption{Relative deviation of the analytic power spectrum (\ref{eq:43}) from the numerical, non-linear result from \cite{2003MNRAS.341.1311S} at $a = 1$.}
\label{fig:3}
\end{figure}

The figure shows that our analytic equation (\ref{eq:43}) reproduces the fully non-linear density-fluctuation power spectrum obtained from numerical simulations \cite{2003MNRAS.341.1311S} very well. The relative difference between $\bar{\mathcal{P}}(k,t)$ and the numerical result for $a = 1$ is on average $\sim15\%$ up to $k = 10\,h\,\mathrm{Mpc}^{-1}$; cf.\ Fig.~\ref{fig:3}. Since, as shown in Fig.~\ref{fig:1}, the momentum-diffusion term dominates the interaction term on small scales, the characteristic non-linear deformation of the power spectrum is shaped by the initial momentum correlations and not by the particle interactions.

\section{Summary}

Starting from our kinetic field theory (KFT) for cosmic structure formation as developed in earlier papers, we have shown here how the interaction term in the generating functional of KFT can be evaluated applying the Born approximation to the particle trajectories. We average this interaction term over the correlated particle distribution and integrate it over time. Inserting it into the two-point density function derived from the generating functional, we arrive at a closed analytic, non-perturbative expression for the non-linear cosmic density-fluctuation power spectrum evolved to the present time. Our expression is free of adjustable parameters and reproduces numerical results with an average accuracy of $\sim15\%$ to wave numbers $k = 10\,h\,\mathrm{Mpc}^{-1}$.

Our derivation shows that the particle interactions only partially compensate the momentum-diffusion term inevitably occuring due to momentum fluctuations in the particle ensemble. The characteristic deformation of the non-linear power spectrum compared to the input power spectrum is dominated by the initial correlations of the particle momenta.

Our main result (\ref{eq:43}) can now straightforwardly be generalized to different cosmologies, initial conditions, or gravitational laws. Deriving higher-order spectra such as the cosmic density bi- and trispectra should also be equally straightforward. Deviations from the Born approximation can now be studied by a low-order perturbative approach relative to the Born approximation, which can safely be expected to converge quickly.

\begin{acknowledgments}
  This work was supported in part by the Transregional Collaborative Research Centre TR~33 ``The Dark Universe'' of the German Science Foundation and by the Excellence Initiative of the German Federal and State Governments. We are grateful for substantial support, many constructive and helpful discussions in particular to L.\ Amendola, J.\ Berges, R.\ Durrer, J.\ Pawlowski, M.\ Salmhofer, B.\ M.\ Sch\"afer, G.\ Starkman, C.\ Wetterich and A.\ Wipf.
\end{acknowledgments}

\bibliographystyle{aipnum4-1}
\bibliography{../bibliography/main}

\end{document}